# Search and Study of Low-Mass Scalar Mesons in the Reaction $np \to np\pi^+\pi^-$ at the Impulse of Neutron Beam $P_n = 5,20 \pm 0,12 GeV/c$

Yu.A.Troyan[a*], A.P.Jerusalimov[a], A.V.Beyaev[a], A.Yu.Troyan[a*], E.B.Plekhanov[a], S.G.Arakelian[b]

[a] – VBLHEP JINR Dubna Russia; [b] – P.N.Lebedev Physical Institute, Moscow, Russia; [*] – atroyan@jinr.ru

**Abstract**

Using irradiation of a 1-meter Hydrogen Bubble Chamber (LHE JINR) by a quasimonochromatic neutron beam with impulse $P_n = 5,20 \pm 0,12 GeV/c$ produced after acceleration of deuterons in the synchrophasotron (LHE JINR) 2560 events of reaction $np \to np\pi^+\pi^-$ were selected. After supplementary sorting out the events where a secondary proton flies forward in the general c.m.s. of reaction ($\cos\Theta_p^* > 0$) in the effective mass spectrum of $\pi^+\pi^-$ - combinations, there were nine peculiarities found out at masses 350, 408, 489, 579, 676, 762, 878, 1036, 1170 $MeV/c^2$ with experimental widths not more than several tens of $MeV/c^2$.

The resonance effects grow too much after using an additional criterion – balance of the sum of the longitudinal impulse of $\pi^+$ and $\pi^-$-mesons in the general c.m.s. of the reaction. Under these conditions the values of a standard derivation from the background are 5 ÷ 7.5.

The direct measurement of the spin of resonances was carried out. Also other quantum numbers were obtained. All of these peculiarities have a similar set of quantum numbers $I^G\left(J^{PC}\right) = 0^+\left(0^{++}\right)$.

This way the sequence of scalar-izoscalar resonances $f_0(\sigma_0)$ with masses in the range of $M \le 1200 MeV/c^2$ are to be seen. There is no information about these resonances in the literature.

The phenomenological dependence for the resonance mass from its number was found. This dependence covered not only resonances shown in this paper but also all those which are present in PDG tables with quantum numbers $f_0(\sigma_0)$ - mesons.

## Introduction

This paper is devoted to the search and study of low-mass $\left(M \le 1.2 GeV/c^2\right)$ resonances in the $\pi^+\pi^-$-system. Their existence can clarify properties of scalar mesons (so called $\sigma_0$-mesons), whose studying is important to understand the mechanism of the chyral symmetry realization (and the mechanism of hadron deconfinement related with it), as well as to understand the attractive part of the nucleon-nucleon potential [1].

In paper [2] it was noted that the light scalar mesons constitute the Higgs sector of strong interactions that ensure finite masses of all kinds of light hadrons (with simultaneous disappearance of $\pi$ – meson mass).

To clarify these fundamental conceptions, it is necessary to carry out a thorough study of properties of $\sigma_0$-mesons and also investigate their structure.

There are hundreds of articles devoted to these problems, and we have no possibility to quote all of them. We send a reader to the review in PDG [3]. However, the review [3] of some theoretical works concerns the quark-gluon picture of the scalar mesons structure. There $\sigma_0$-mesons are considered as a construction of 2 or 4 quarks, as gluebols, and as constructed of quarks into diquarks, etc.

But there are some other conceptions of this problem, among them – the predictions of series and properties of resonances from the point of view of multi-dimensional space [4], predictions of series of resonances, obtained from formulae of quasi-classic quantization [5], predictions of resonance series, based on the McGregor model [6], the search for properties of instant vacuum [7].

Such a large number of theoretical conceptions confirms that the understanding of properties and structure of $\sigma_0$-mesons does not exist yet.

It may be explained, first of all, by the absence of experimental data in the region of masses less than 1 GeV/c$^2$. The problem becomes more important due to: statistically abundant data obtained at the HADES spectrometer [8]; investigations of the mixed phase planned at the JINR [9]; some similar experiments planned in other accelerators in the world.

The low-mass $\sigma_0$-mesons can become a powerful tool to study a new state of the matter. Some predictions about varying of the $\sigma_0$-meson properties under intermediate conditions, have been obtained [10].

## 1. Reaction $np \to np\pi^+\pi^-$

This paper continues a series of our investigations, devoted to the study of low-mass $\left(M \leq 1.2GeV/c^2\right)$ resonances in the $\pi^+\pi^-$ - system in the reaction $np \to np\pi^+\pi^-$ at $P_n = 5,20 \pm 0,12GeV/c$ [11,12].

The study was carried out using the data obtained in the exposure of the 1-m. HBC (LHE JINR) to a quasimonochromatic neutron beam with $\Delta P_n / P_n \approx 2.5\%, \Delta\Omega_n \approx 10^{-7}$ *sterad.* due to the acceleration of deuterons by synchrophasotron of LHE [13].

The accuracies of momentum of secondary charged particles from the reaction are: $\delta P \approx 2\%$ for protons and $\delta P \approx 3\%$ - for mesons. The angular accuracy was $\leq 0.5°$.

The channels of the reactions were separated by the standard $\chi^2$–method taking into account the corresponding coupling equations [14-16].

In this work in comparison with the previous ones we have made a more strict selection of the events by the value of $\chi^2$ for using events: $\chi^2 < 1,5$. Some cut of the missing mass reconstructed after the fit-procedure was performed.

The $\chi^2$ – distribution is shown in Fig.1a. The theoretical curve for $\chi^2$ – distribution with one degree of freedom is shown in the same Figure. In Fig.1b the missing mass distribution in the selected events reaction $np \to np\pi^+\pi^-$ is illustrated. One can see that the distribution has a maximum at the missing mass equal to the neutron mass with accuracy of 0.1 MeV/c$^2$ and is symmetric relately the neutron mass. The width at the half-height is 20 MeV/c$^2$. For a better purity of the data a small number of events with the missing masses out of the gray range was excluded.

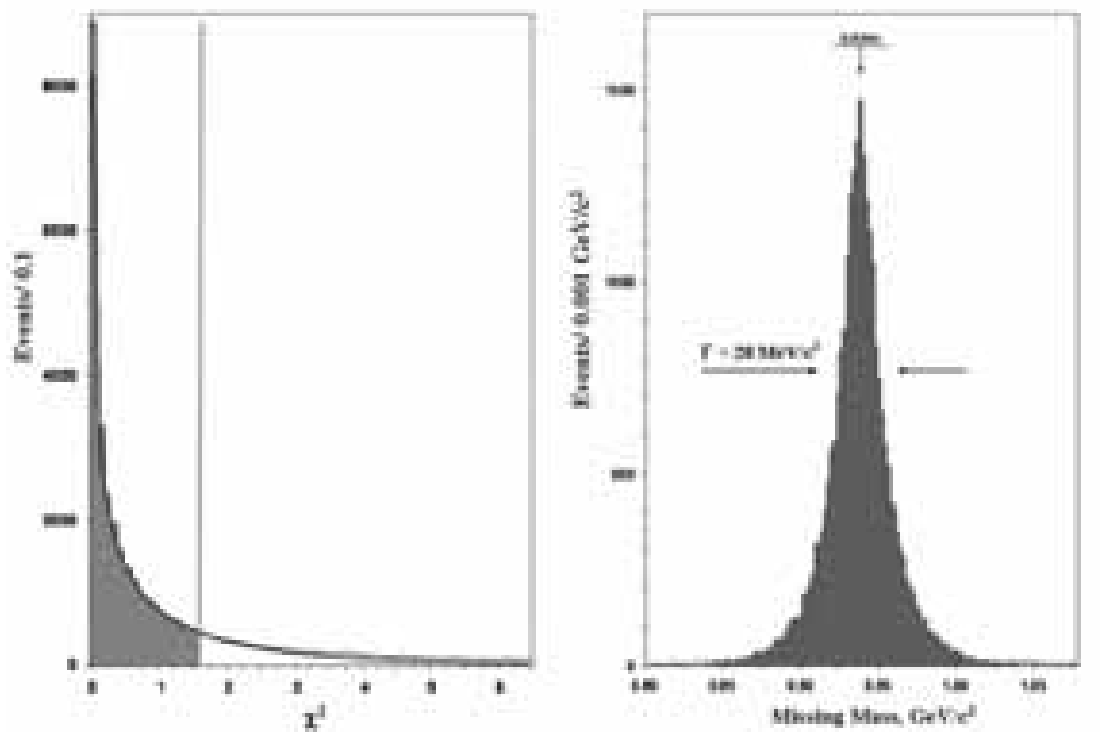

**Figure 1**

**a)** The histogram - the experimental $\chi^2$ – distribution for the events from the reaction $np \to np\pi^+\pi^-$; line - the theoretical curve for $\chi^2$ – distribution with one degree of freedom

**b)** The missing mass distribution in the selected reaction $np \to np\pi^+\pi^-$ with $\chi^2 < 1,5$

**Figure 1 a) b)**

As a result, 25650 events of the $np \to np\pi^+\pi^-$ reaction were selected under $4\pi$-geometry after application of all the cuts. Note that an admixture of other reactions is practically absent.

Earlier we have already studied the reaction $np \to np\pi^+\pi^-$ [17], and OPE-exchange with a dominating charged $\pi$-meson part has been shown as the main mechanism of this reaction. It leads to a plentiful production of $\Delta^{++}$ and $\Delta^-$-resonances (up to 70% of the reaction cross section). This OPE-exchange gives the main part into the events with a neutron flying into the forward hemisphere in the general c.m.s. of the reaction.

Figure.2 demonstrated the effective mass distribution of $\pi^+\pi^-$-combinations for the events with $\cos\Theta_n^* > 0$. There are no such effects on this distribution. The superposition of Legendre polynomials up to the 9$^{th}$ power (solid line) describes this experimental distribution with $\overline{\chi^2} = 1.02 \pm 0.15$ and $\sqrt{D} = 1.51 \pm 0.11$

The background curve (dashed line) calculated by means of the OPER model (One Pion Reggezied Exchange) is also imposed on the distribution in Fig.2. The OPER model contains the processes of $N^*$ and $\Delta^*$-resonances production and elastic $\pi\pi \to \pi\pi$ scattering. In addition to the description of the reaction $np \to np\pi^+\pi^-$, the processes of the diffraction production of $N^*_{1440}$, $N^*_{1520}$ and $N^*_{1680}$ -resonances are used. This curve describes the distribution with $\overline{\chi^2} = 1,07 \pm 0,15$ and $\sqrt{D} = 1,60 \pm 0,11$.

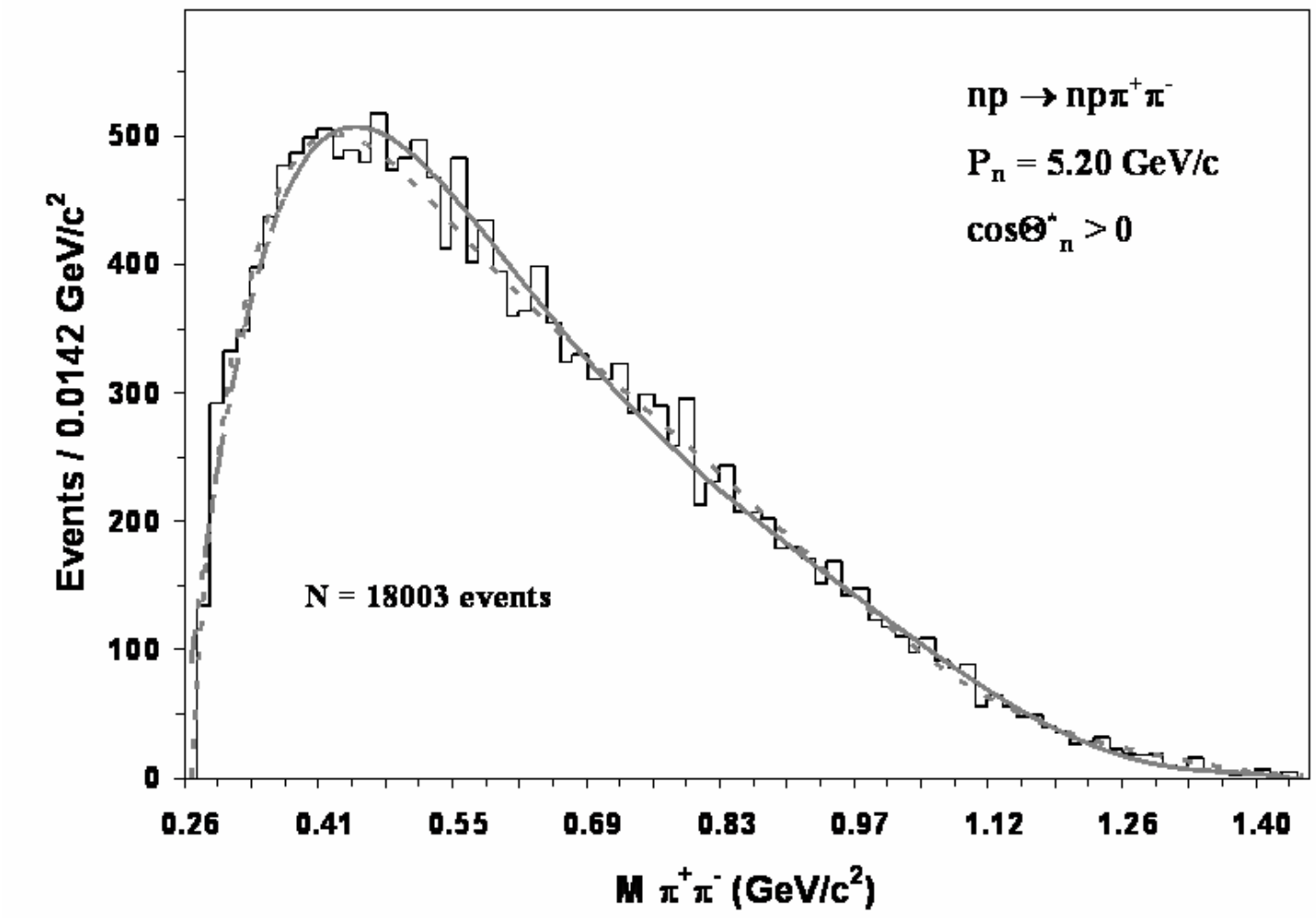

**Figure 2** The effective mass distribution of $\pi^+\pi^-$-combinations for the events with $\cos\Theta^*_n > 0$.
The solid line - the superposition of Legendre polynomials up to 9th power
The dashed line - the background curve calculated by means of the OPER model

Therefore, it seems reasonable to study the resonances in the $\pi^+\pi^-$-system of the reaction $np \to np\pi^+\pi^-$ selecting the events at the condition that $\cos\Theta^*_p > 0$. The total contribution of the $\Delta^{++}$ and $\Delta^-$- resonances is not more than 17% for these events (calculated by OPER model), and the background from the resonance decays decreases greatly.

Figure.3 shows the effective mass distribution of $\pi^+\pi^-$-combinations for the events with a secondary proton flying into the forward hemisphere in the general c.m.s. of the reaction ($\cos\Theta^*_p > 0$). The number of events with $\cos\Theta^*_p > 0$ is equal to 7647.

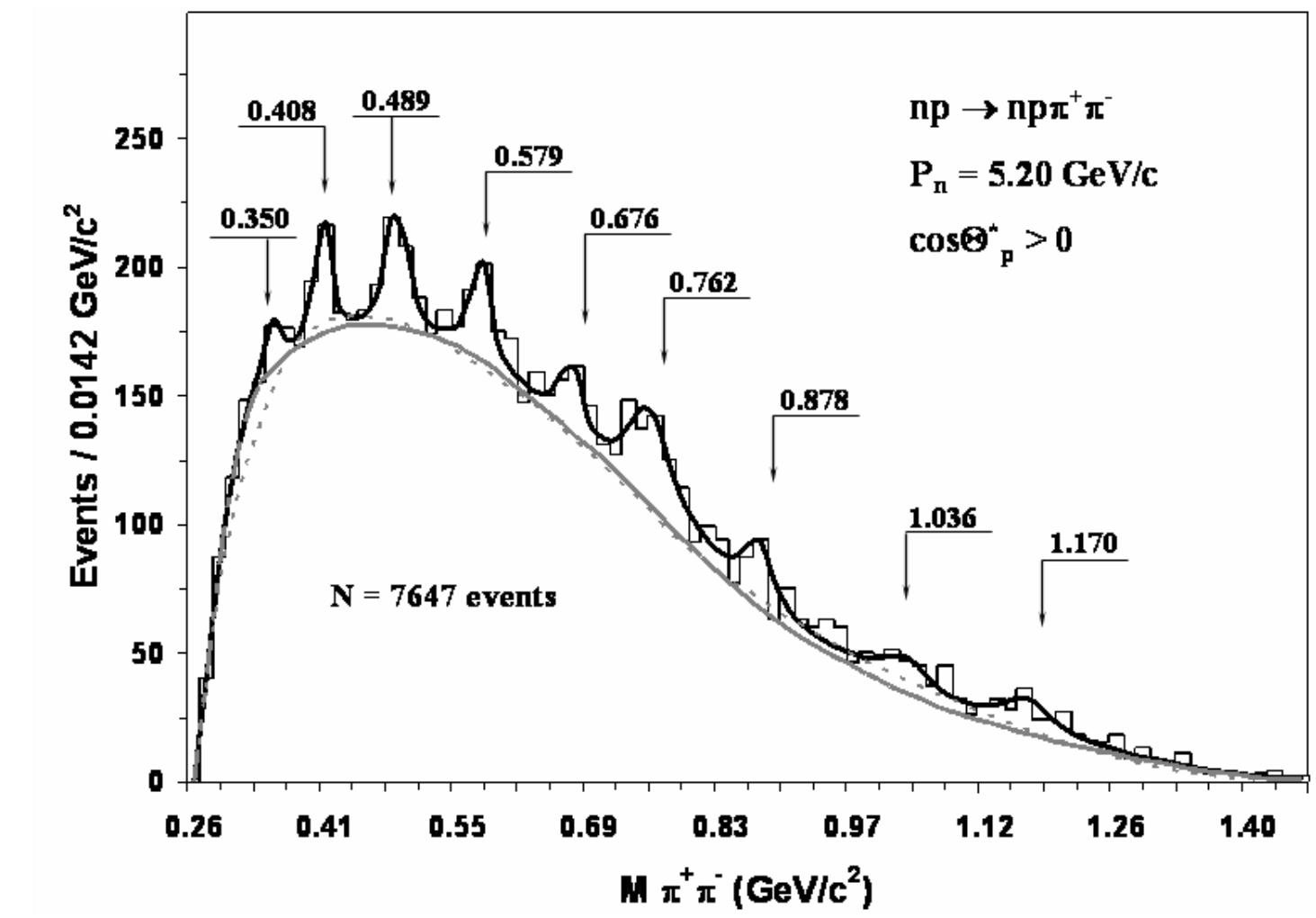

**Figure 3** The effective mass distribution of $\pi^+\pi^-$-combinations for the events with $\cos\Theta^*_p > 0$
The solid line – the approximation by an incoherent sum of the polynomial background and by 9 resonance curves taken in the Breight-Wigner form
The gray line - the background curve of the Legendre polynomials (up to the 9th power)
The dashed line - the background curve calculated by means of the OPER model
The experimental values of the resonance masses are shown with arrows.

The big difference between Fig.2 and Fig.3 is seen.

The mass distribution in Fig.3 is approximated by an incoherent sum of the background curve taken in the form of a superposition of Legendre polynomials up to 9th power (gray line) and by 9 resonance curves taken in the Breight-Wigner form. The experimental values of the resonance masses (obtained by fitting procedure) are shown with arrows.

The requirements to the background curve are as follows:

- the errors of the coefficients must not be more than 50% for each term of the polynomial;
- the polynomial must describe the experimental distribution after "deletion" of the resonance regions with $\overline{\chi^2} = 1.0$ and $\sqrt{D} = 1.41$ (the parameters of $\chi^2$ – distribution with one degree of freedom).

The values $\overline{\chi^2}$ and $\sqrt{D}$ for the background curve (red line) of the distribution in Fig.3 (with resonance regions excluded) are equal to $\overline{\chi^2} = 0.97 \pm 0.24$ and $\sqrt{D} = 1,36 \pm 0,17$. The contribution of the background to this distribution is 89%.

The same values for the background curve normalized to 100% of the events in the plot (with resonance regions included) are equal to $\overline{\chi^2} = 1.26 \pm 0.15$ and $\sqrt{D} = 1,59 \pm 0,11$ (Confidence Level 9%).

The background curve (dashed line) calculated by means of OPER model is also imposed on the distribution in Fig.3. This curve describes the distribution with $\overline{\chi^2} = 0,95 \pm 0,24$ and $\sqrt{D} = 1,45 \pm 0,17$ and, normalized to 100% of the events in the plot (with resonance regions included) with $\overline{\chi^2} = 1,24 \pm 0,15$ and $\sqrt{D} = 1,87 \pm 0,11$ (Confidence Level 11%).

The results of approximation are given in Table 1 below.

**Table 1**

| $M_{Res} \pm \Delta M_{Res}$, MeV/c$^2$ | $\Gamma_{Res} \pm \Delta\Gamma_{Res}$, MeV/c$^2$ | S.D. | σ μb |
|---|---|---|---|
| 350 ± 3 | 11 ± 8 | 3.5 | 12 ± 6 |
| 408 ± 3 | 11 ± 8 | 3.5 | 12 ± 6 |
| 489 ± 3 | 20 ± 10 | 4.0 | 20 ± 8 |
| 579 ± 5 | 17 ± 14 | 3.8 | 18 ± 8 |
| 676 ± 7 | 11 ± 14 | 3.0 | 11 ± 6 |
| 762 ± 11 | 53 ± 33 | 6.1 | 26 ± 8 |
| 878 ± 7 | 30 ± 14 | 3.6 | 11 ± 5 |
| 1036 ± 13 | 61 ± 30 | 5.1 | 15 ± 5 |
| 1170 ± 11 | 65 ± 33 | 5.8 | 11 ± 4 |

*The first column* contains the experimental values of the resonance masses (including errors) obtained in the process of approximation.

*The second column* contains the experimental values of the resonance widths.

The mass resolution function $\Gamma_{res}(M)$ [18] grows with the increasing mass as follows:

$\Gamma_{res}(M) = 4.2\left[\left(M - \sum_{i=1}^{2} m_i\right)/0.1\right] + 2.8$, where:

M – the mass of the resonance, $m_i$ – the rest mass of an *i*-particle including in this resonance, M and $m_i$ are in GeV/c$^2$; coefficients 4.2 and 2.8 are in MeV/c$^2$.

The true width of the resonance is obtained by the following formula: $\Gamma_{Res}^{true} = \sqrt{\left(\Gamma_{Res}^{exp}\right)^2 - \left(\Gamma_{res}\right)^2}$.

*The third column* contains the number of standard deviations of the effects above the background: $S.D. = N_{Res.} / \sqrt{N_{back.}}$.

*The fourth column* contains the production cross sections for the corresponding resonances.

For the cross sections errors, we have taken into account the cross section error for the reaction $np \to np\pi^+\pi^-$ at $P_n = 5.20\,GeV/c$ ($\sigma_{np\to np\pi^+\pi^-} = (6.22 \pm 0.28)mb$) [12].

## 2. Spin and Isotopic Spin of the Resonances

We have tried to estimate the values of spins of the observed resonances in the $\pi^+\pi^-$ - system.

To do this, we have constructed the distributions of emission angles of $\pi^+$ - meson from the resonance decay (events from the corresponding resonance mass region) with respect to the direction of the resonance emission in the general c.m.s. of the reaction. All the values are transformed to the resonance rest system (helicity coordinate system) [16]. The backgrounds are constructed by using the events at the left and right of the corresponding resonance band and subtracted using the weight in proportion to the contribution of the background into the resonance region. The result distributions are described by a sum of the even-power Legendre polynomial with the maximum power being equal to 2J, where J is a lower border of the resonance spin. The authors are grateful to Dr. V. L. Lyuboshitz for the arrangement of the corresponding formulae. All distributions are isotropic, that was showed in our previous works [11, 12]. Therefore, the most probable spin values for these resonances are: $J \geq 0$.

We have not observed the corresponding resonances either in the $\pi^-\pi^0$ - system of the $np \rightarrow pp\pi^-\pi^0$ reaction [11], or in the $\pi^-\pi^-$ - system of the $np \rightarrow pp\pi^+\pi^-\pi^-$ reaction, which also have been studied by us [18].

Therefore, it can be affirmed that all the resonances observed by us have the quantum numbers $I^G\left(J^{PC}\right)=0^+\left(0^{++}\right)$ and may be identified as $\sigma_0$ - mesons.

### 3. Intensification of the effect

To intensify the effect there, an additional criterion to select of the events was introduced – namely on the variable $X^*_{\pi^+\pi^-}=\left(P^*_{\|\pi^+}+P^*_{\|\pi^-}\right)/P^*_{\pi\ \max}$, where: $P^*_{\|\pi^{+(-)}}$ - the experimental value of the longitudinal component of $\pi^+\left(\pi^-\right)$ in the general c.m.s., $P^*_{\pi\ \max}$ - the maximum possible value of π-meson momentum for the given event in the general c.m.s.

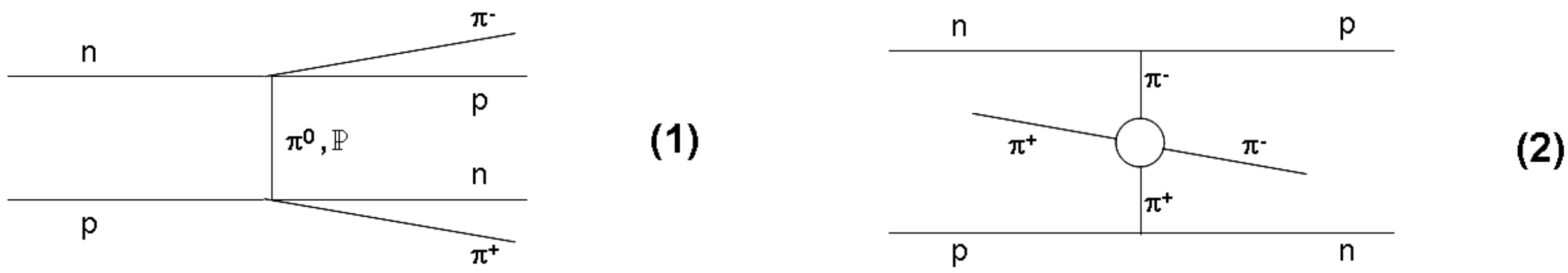

**Figure. 4** The examples of the diagrams 1) - OPE where the exchange is $\pi^0$ – meson (Pomeron)
2) – the process of $\pi^+\pi^-$ - scattering

Since the application of the criterion $\cos\Theta^*_p > 0$, the main background arises due to the OPE diagrams, where the exchange is $\pi^0$ – meson (diagram of type (1) in Fig.4) and due to the process of diffraction production (Pomeron exchange): the process $\pi^- p \rightarrow \pi^0 n$ (charge exchange process) plays the basic part in these diagrams. There are no peculiarities in the cross section of this process except the maximum at $M_{p\pi^-}=1236 MeV/c^2$.

The diagrams of $\pi^+\pi^-$ - scattering (diagram of type (2) in Fig.4) begin to play a much more important part. This diagram was earlier vastly masked by diagrams of the charged meson exchange. In diagrams of type (2) the $\pi^+$ and the $\pi^-$ are equitable, therefore it is possible to expect that the effect will arise in $\pi^+\pi^-$ - scattering, and the $\pi^+\pi^-$ - system will be close to the symmetric one (in $X^*_{\pi^+\pi^-}$ variable) relatively to $X^*_{\pi^+\pi^-}\approx 0$.

Let us consider the application of $X^*_{\pi^+\pi^-}$ variable to intensify the resonance effect at the mass of 489 MeV/$c^2$. The similar procedure was used to study other effects.

Fig. 5a presents the distributions of $X^*_{\pi^+\pi^-}$. The solid line shows the distribution of the events for the mass region of $M_{\pi^+\pi^-}=489 MeV/c^2$. The dashed line shows the distribution of the events for masses from the left and right of this mass region (background events). The resonance and background regions of masses are shown in Fig.5b. One can see that the resonance events get to some regions of $X^*_{\pi^+\pi^-}$ with a more probability than the background one. It is possible significantly to decrease a part of the background to the resonance due to selection of the bands on $X^*_{\pi^+\pi^-}$ where the probability for the resonance events is bigger than for the background one.

The distribution of the $\pi^+\pi^-$ effective masses is shown in Fig. 6a for the events when the value of $X^*_{\pi^+\pi^-}$ reaches the regions marked in Fig. 5a (filled bands). One can see a sharp increase of the effect at the mass of $M_{\pi^+\pi^-}=489 MeV/c^2$. Then its statistical significance is: S.D.=6.

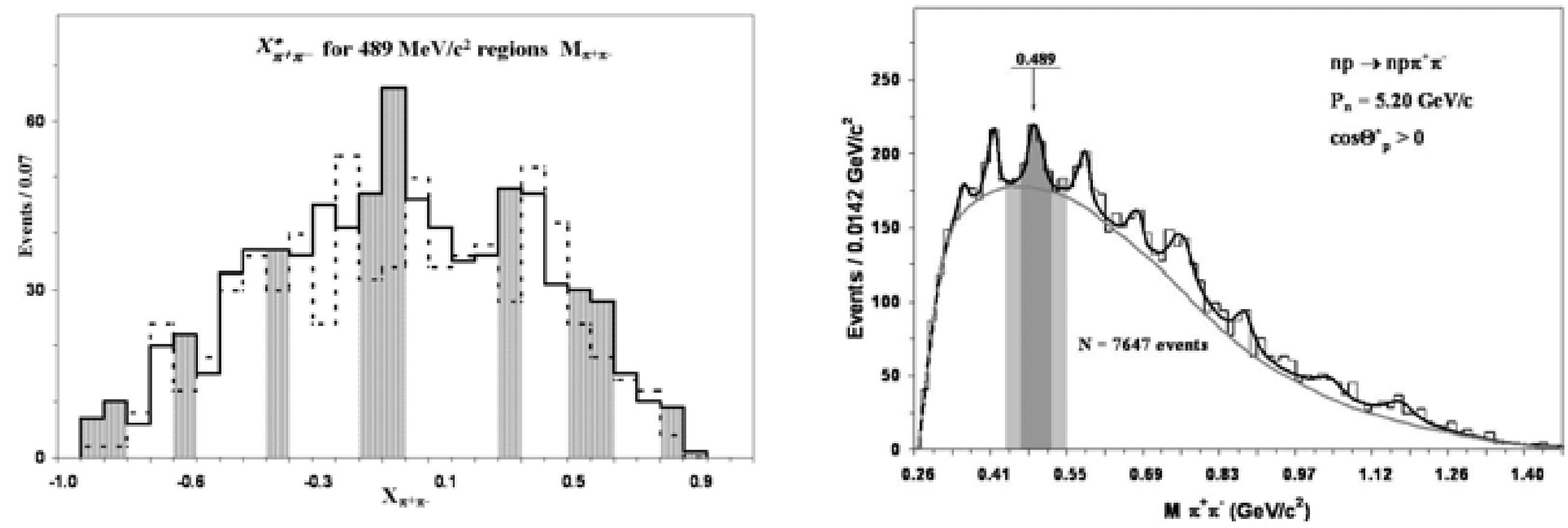

**Figure 5** **a)** **b)**

**a)** The distributions on $X^*_{\pi^+\pi^-}$. The solid line – the distribution of the events for the mass region of the resonance at the mass of 489 MeV/c$^2$. The dashed line – the distribution of the events for masses from the left and right of the resonance at the mass of 489 MeV/c$^2$ (background events). The gray bands – the selected regions of $X^*_{\pi^+\pi^-}$

**b)** The resonance and background regions for the resonance at the mass of 489 MeV/c$^2$
The gray bands – the background regions. The dark-gray bands – the resonance region

The distribution of the $\pi^+\pi^-$ effective masses is shown in Fig.6b for the events when the value of $X^*_{\pi^+\pi^-}$ is outside the regions marked in Fig.5a. One can see the total absence of the effect at the mass of $M_{\pi^+\pi^-} = 489 MeV/c^2$.

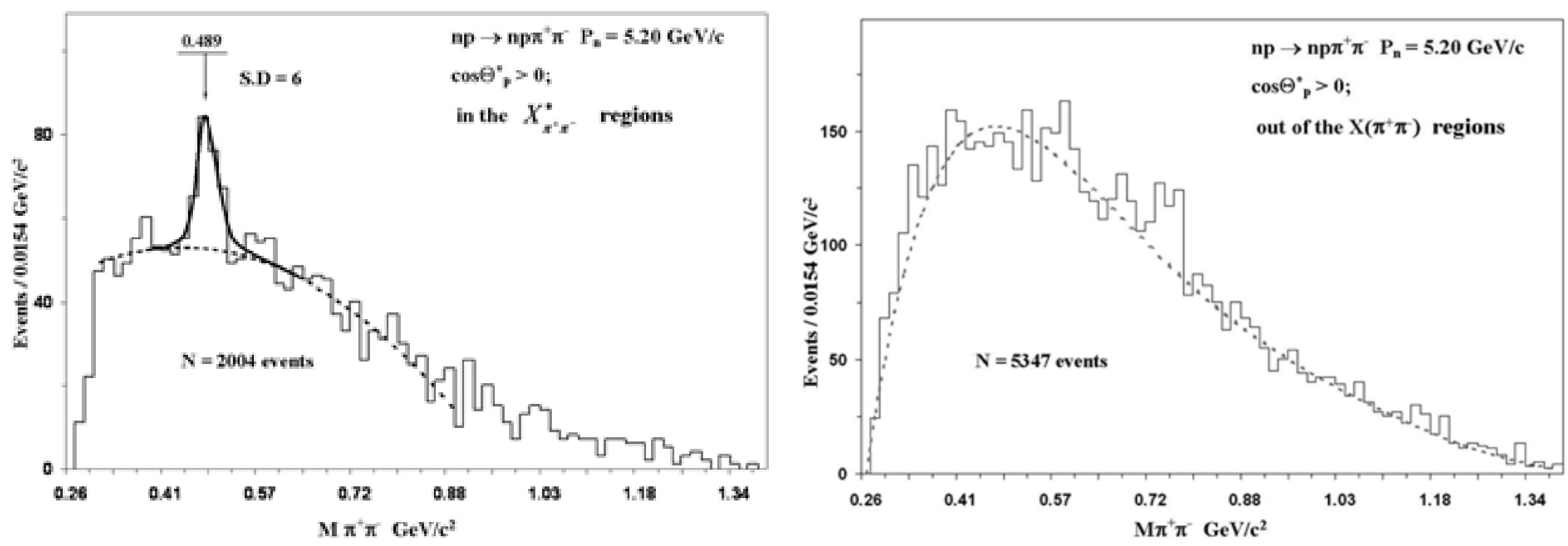

**Figure 6** **a)** **b)**

**c)** The distribution of the $\pi^+\pi^-$ effective masses for the events when the value of $X^*_{\pi^+\pi^-}$ get to the regions marked in Fig. 5a. The dashed line – the polynomial background. The solid line – the resonance curve taken in the Breight-Wigner form

**d)** The distribution of the $\pi^+\pi^-$ effective masses for the events when the value of $X^*_{\pi^+\pi^-}$ is outside the regions marked in Fig.5a. Dashed line – the curve of the polynomial approximation

The same analysis of the corresponding distributions of $X^*_{\pi^+\pi^-}$ was performed for all the effects observed by us.

Fig.7 shows the regions of samples on $X^*_{\pi^+\pi^-}$ for the various resonance mass. Some of the selected regions are overlapping for different resonances. This observation is taken into account at the construction of the background curves for the distributions of the resonance effective masses. The backgrounds are constructed by a part of the points which are not related with other resonances by overlapping on the Fig.7 ($X^*_{\pi^+\pi^-}$ regions).

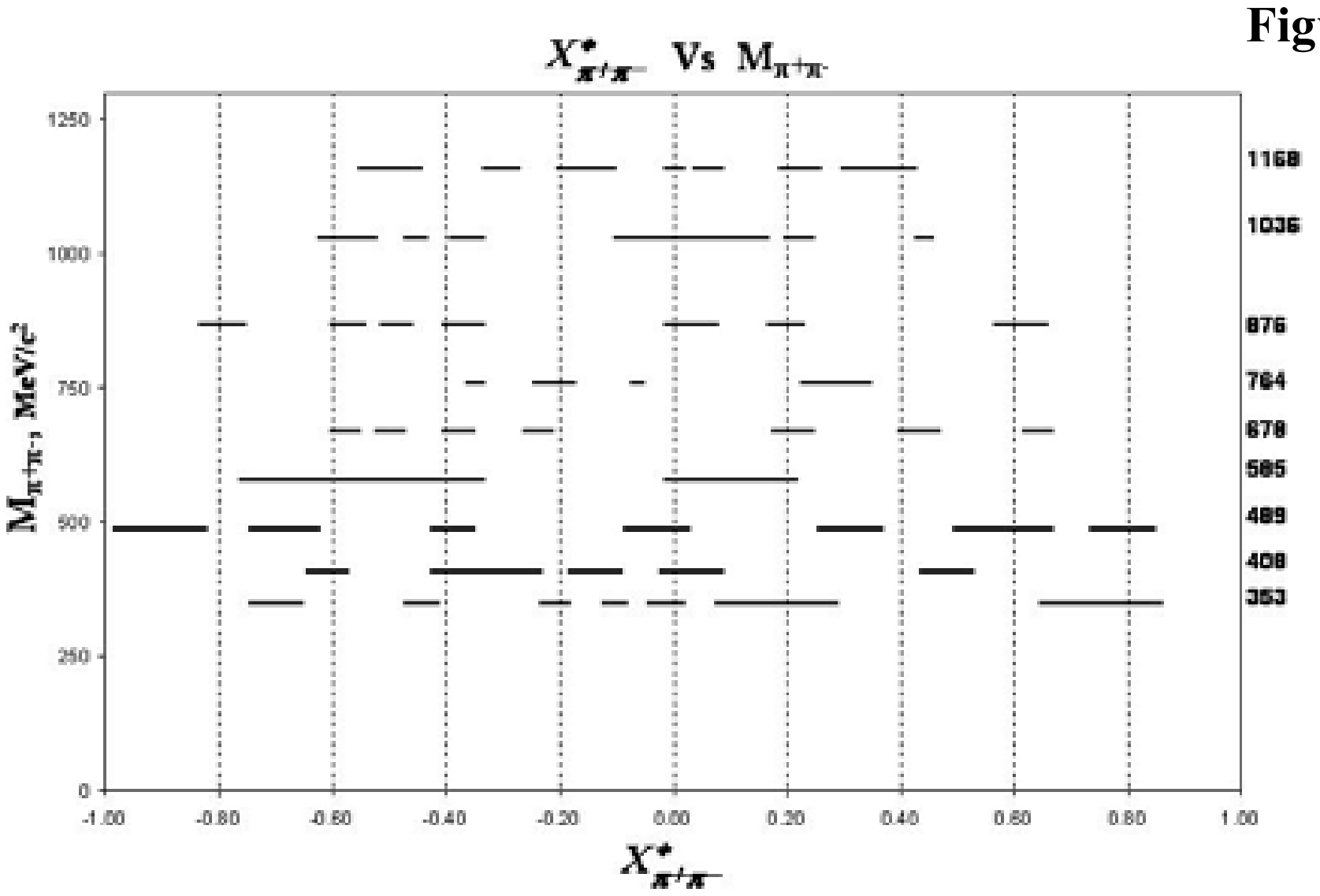

**Figure 7**
The regions of samples on $X^*_{\pi^+\pi^-}$ for the various resonance mass. The values of the resonance mass are to the right of the picture.

Fig.8 presents the distributions of $M_{\pi^+\pi^-}$ (since use the criterion $X^*_{\pi^+\pi^-}$) for all the resonances observed by us. The background and resonance (Breight-Wigner) curves are shown in the same place.

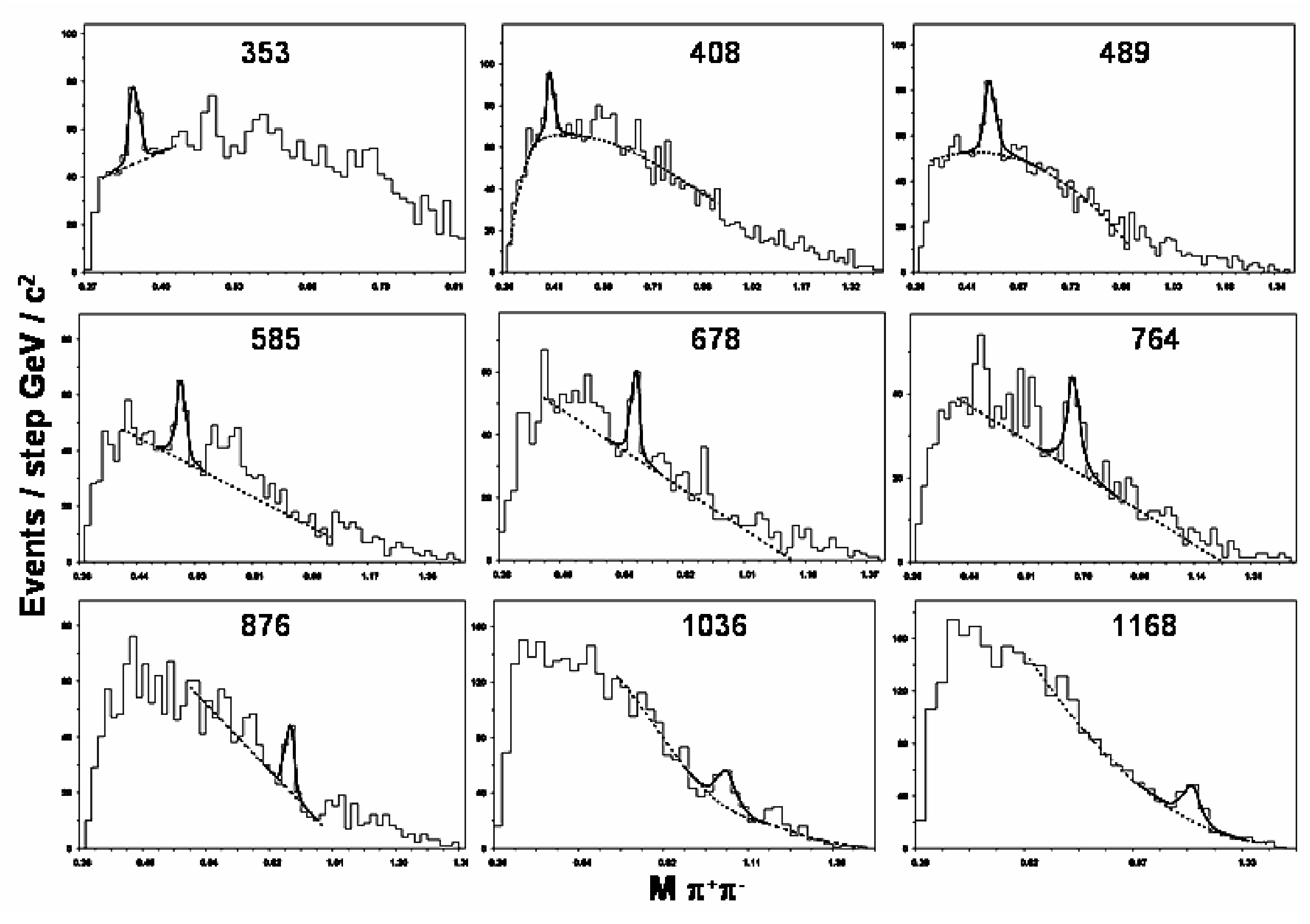

**Figure 8** The distributions of $M_{\pi^+\pi^-}$ for all resonances observed since criterion $X^*_{\pi^+\pi^-}$ has been introduced.
The mass values are in the pictures. The dashed line - the polynomial background. The solid line – the resonance curve taken in the Breight-Wigner form.

Table 2 shows the fitted values of the masses and widths for the resonances as well as the calculated values of the standard deviations above the background. One can see the values of the resonances masses and widths to be close to the values shown in Table 1, and the values of S.D. significantly increase and are more than 5 everywhere. In the result the use of the $X^{*}_{\pi^+\pi^-}$ criterion has strongly decreased the level of the background for each effect but has not distorted the resonance characteristics.

| $M_{Res} \pm \Delta M_{Res}$, MeV/c$^2$ | $\Gamma_{Res} \pm \Delta\Gamma_{Res}$, MeV/c$^2$ | S.D. |
|---|---|---|
| 350 ± 3 | 11 ± 8 | 3.5 |
| 408 ± 3 | 11 ± 8 | 3.5 |
| 489 ± 3 | 20 ± 10 | 4.0 |
| 579 ± 5 | 17 ± 14 | 3.8 |
| 676 ± 7 | 11 ± 14 | 3.0 |
| 762 ± 11 | 53 ± 33 | 6.1 |
| 878 ± 7 | 30 ± 14 | 3.6 |
| 1036 ± 13 | 61 ± 30 | 5.1 |
| 1170 ± 11 | 65 ± 33 | 5.8 |

**Table 2** The fitted values of the masses and widths for resonances shown in Fig.8
*The first column* contains the experimental values of the resonance masses (including errors) obtained in the process of approximation.
*The second column* contains the experimental values of the resonance widths.
*The third column* contains the number of standard deviations of the effects above the background

Fig.9 shows the distributions of the emission angles of $\pi^+$ - mesons from the resonance decay (in the helicity coordinate system). These distributions are obtained subtracting the corresponding background curves for intense effects using all the criteria mentioned above. One can see that all the distributions are isotropic. This implies that $J \geq 0$ for all the observed peculiarities.

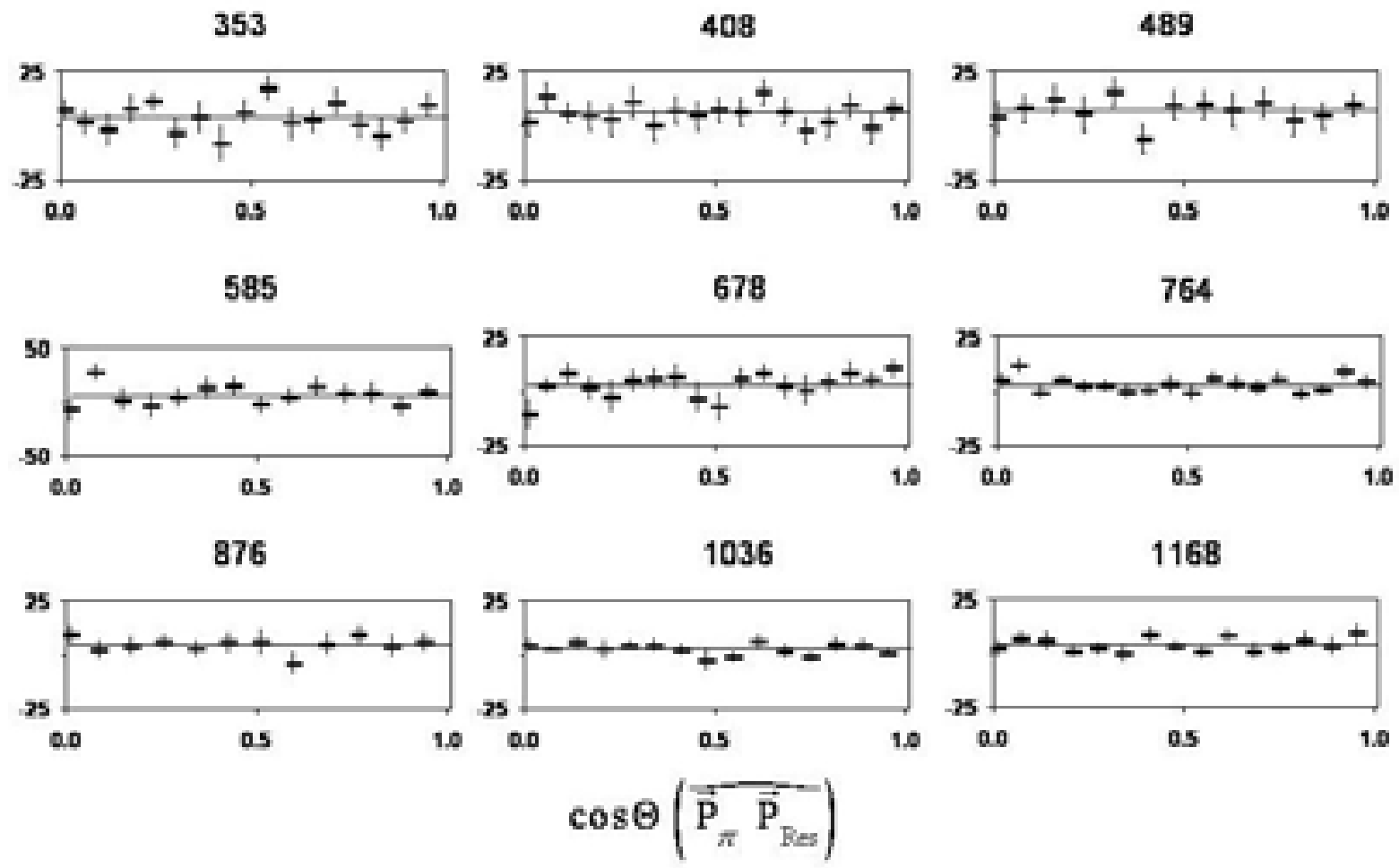

**Figure 9** The distributions of the emission angles of $\pi^+$ - mesons from the resonance decay in the helicity coordinate system for intense effects using all the criteria mentioned for the intensed effects.
The values of the resonance mass are shown.
The solid lines demonstrate the isotropic distributions.

## 4. Comparison with other experiments

There are too few experiments that can be compared with our data.

First, it is the experiment where the so-called ABC-effect was observed [19]. It is the peak at the mass of $M = 350 MeV/c^2$ that was found in various nuclear reactions. We have observed [20] a similar effect in the reaction $np \to d\pi^+\pi^-$ at $P_n = (1.73 \pm 0.04) GeV/c$ in the system of $\pi^+\pi^-$ at the mass of $M_{\pi^+\pi^-} \approx 400 MeV/c^2$.

The resonance peculiarity was found in $dC$ - interactions at the momentum of 2.75MeV/c per nucleon in the effective mass spectrum of $2\gamma$ at the mass of $M_{\gamma\gamma} = 360 MeV/c^2$ [21].

The effects there were found at the effective masses of $2\gamma$ $M_{\gamma\gamma} = 350 MeV/c^2$ in the 2-m propane bubble chamber. The peaks in the momentum distribution of γ's can be explained under the assumption that the resonances produced at the masses of $M_{\gamma\gamma} = 350 MeV/c^2$ were decaying into 2γ's [22].

Recently the scalar pole was found by PWA of $K_s^0$ – mesons into 2 π – mesons and 4 leptons at the mass of 489 MeV/c$^2$ [23]. But the width of the resonance is very large - $\Gamma/2 = 264 MeV/c^2$ . The pole in the decay $D^+ \to \pi^+\pi^+\pi^+$ was also observed in the system of $\pi^+\pi^-$ at the mass of $M_{\pi^+\pi^-} = 478 MeV/c^2$ [24].

Finally, we show the table from PDG [3] with results of K-matrix analysis of a number of experiments on search for low-mass $\pi^+\pi^-$ peculiarities. Note as well that this table also contains (marked "no PWA") our data [12] concerning the resonance at the mass of $M_{\pi^+\pi^-} = 762 MeV/c^2$ obtained during the direct study of the mass spectrum. The data of K-matrix analysis are in a good agreement with our observations except the resonances widths.

## 5. Phenomenological description of the scalar resonances series

We have undertaken an attempt to describe phenomenologically the scalar $\left(f_0\left(\sigma_0\right)\right)$ resonances taking into account both the resonances observed by us and the data presented in PDG.

The analysis of all the data has shown that the distance between the resonances varies in a rather complicated manner. It leads to an idea of the existence of several trajectories describing the known masses of the resonances.

The following form was taken for the approximation:

$$\boldsymbol{M_n = M_0 + x\frac{n(n+1)}{2}} \qquad \textbf{(1)}.$$

where, $M_0$ – initial excitation of the trajectory; *x* – parameter of the excitation, *n* – number of the resonances on the trajectory. The term of the form $\frac{n(n+1)}{2}$ is of the sum of the natural numbers from *1* to *n*.

Therefore the mass of the $n^{\text{th}}$ - resonance can be presented schematically (Fig.10) in the form of the sum of the series excited by force *x*:

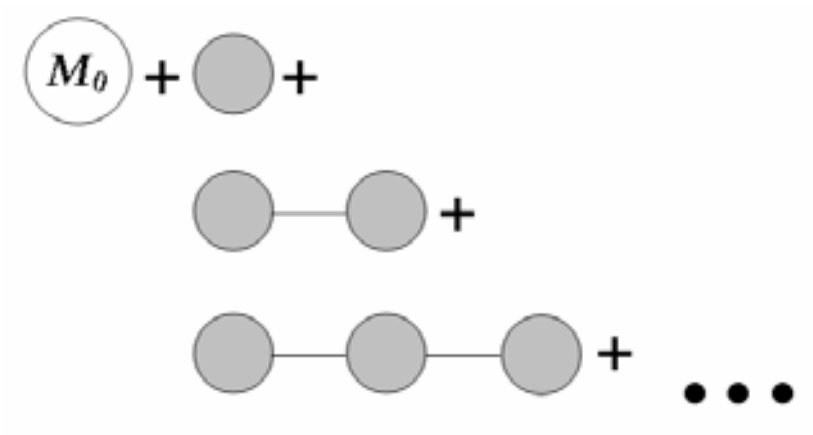

**Figure 10** The schematic presentation the mass of the $n^{\text{th}}$ - resonance in the form of the sum of the series excited by force *x*

Summing up all the series, we get form (1).

We have constructed 4 resonances series for which the values $M_0^i$ and $x^i$ were determined (*i* is a number of series).

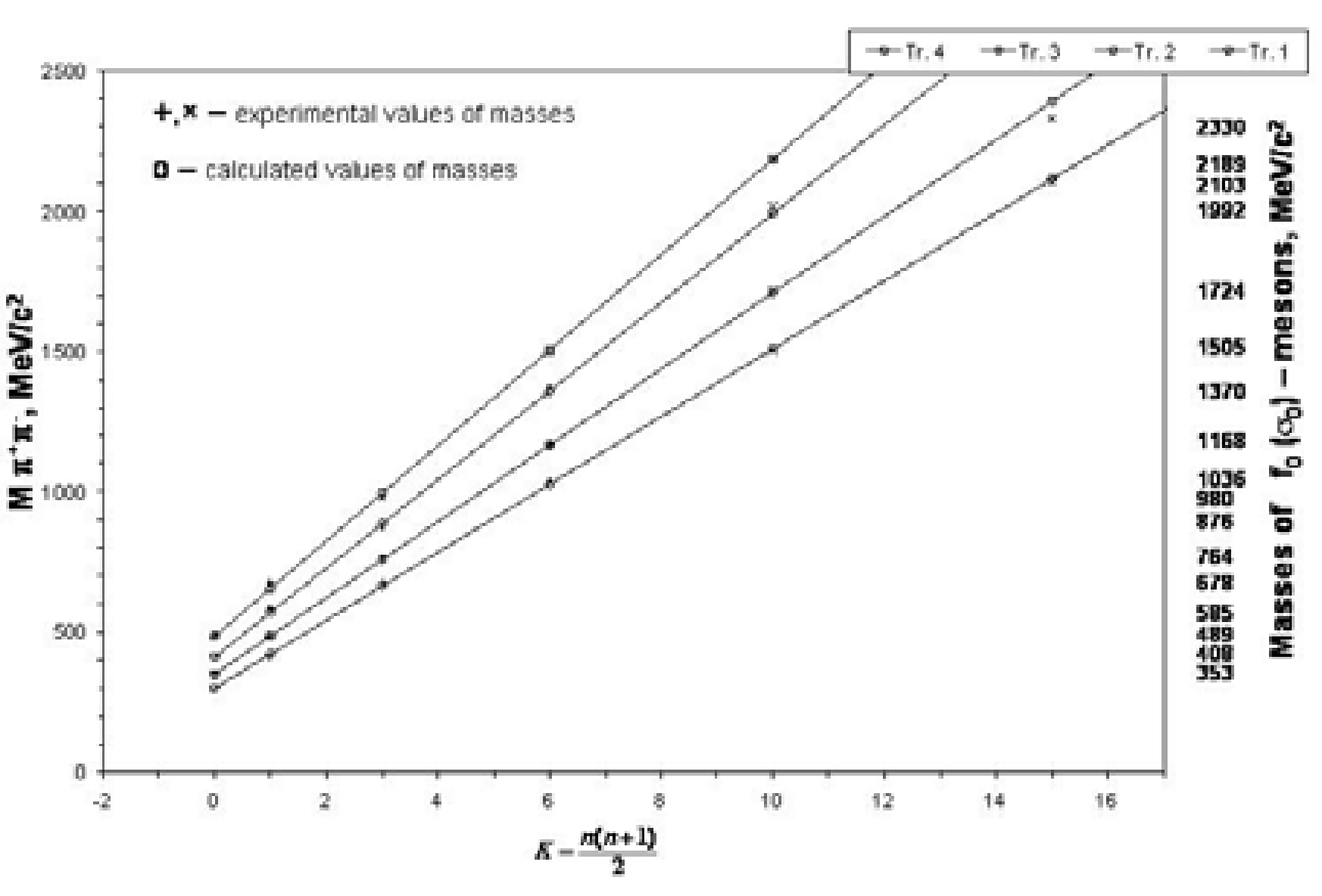

**Figure 11** The trajectories presented in Table 3. The value $K = \frac{n(n+1)}{2}$ is plotted on the abscissa. The masses of the $\pi^+\pi^-$ - combinations are plotted on the ordinate axis. The experimental values of the scalar $\left(f_0\left(\sigma_0\right)\right)$ – mesons are shown in the right side of the Figure. The experimental data from our investigations are pointed as (+) and the data from PDG as (**x**). The values of the experimental errors are less than the sizes of the markers in the Figure.

The 4 constructed trajectories contain all (without exception) the resonances observed by us and the resonances presented in PDG tables. We have considered the masses less than 2400 Mev/c$^2$ because higher the data were very indefinite. These trajectories are shown in Fig.11

Table 3 shows the calculated values of the resonances masses depending on number n for each of the series, the initial excitations $M_0^i$, the excitations of clusters $x^i$ and experimental values of the masses of scalar resonances both from our experiments and PDG tables.

**Table 3**

| **Trajectory 1** $M_n = 305 + 120\frac{n(n+1)}{2}$ | | | | | |
|---|---|---|---|---|---|
| **n** | **n(n+1)/2** | **$M_n$ calc** | **M** | **±** | **ΔM exp** |
| 0 | 0 | 305 | | | |
| 1 | 1 | 425 | 408 | ± | 5 |
| 2 | 3 | 665 | 678 | ± | 3 |
| 3 | 6 | 1025 | 1036 | ± | 10 |
| 4 | 10 | 1505 | 1505 | ± | 6 |
| 5 | 15 | 2105 | 2103 | ± | 8 |

| **Trajectory 2** $M_n = 350 + 136\frac{n(n+1)}{2}$ | | | | | |
|---|---|---|---|---|---|
| **n** | **n(n+1)/2** | **$M_n$ calc** | **M exp** | **±** | **ΔM exp** |
| 0 | 0 | 350 | 353 | ± | 6 |
| 1 | 1 | 486 | 489 | ± | 4 |
| 2 | 3 | 758 | 764 | ± | 18 |
| 3 | 6 | 1166 | 1168 | ± | 11 |
| 4 | 10 | 1710 | 1724 | ± | 7 |
| 5 | 15 | 2390 | 2330 | ± | 20 |

| **Trajectory 3** $M_n = 420 + 158\frac{n(n+1)}{2}$ | | | | | |
|---|---|---|---|---|---|
| **n** | **n(n+1)/2** | **$M_n$ calc** | **M exp** | **±** | **ΔM exp** |
| 0 | 0 | 420 | 408 | ± | 5 |
| 1 | 1 | 578 | 585 | ± | 6 |
| 2 | 3 | 894 | 876 | ± | 4 |
| 3 | 6 | 1368 | 1370 | ± | |
| 4 | 10 | 2000 | 1992 | ± | 16 |

| **Trajectory 4** $M_n = 485 + 170\frac{n(n+1)}{2}$ | | | | | |
|---|---|---|---|---|---|
| **n** | **n(n+1)/2** | **$M_n$ calc** | **M exp** | **±** | **ΔM exp** |
| 0 | 0 | 485 | 489 | | 4 |
| 1 | 1 | 655 | 678 | ± | 3 |
| 2 | 3 | 995 | 980 | ± | 10 |
| 3 | 6 | 1505 | 1505 | ± | 6 |
| 4 | 10 | 2185 | 2189 | ± | 13 |

**Table 3** The calculated values of the resonances masses depending on the number *n* for each of the series
*The first column* contains number of the resonance on the trajectory
*The second column* contains the values of the n(n+1)/2 for number of the resonance on the trajectory
*The third column* contains the calculated values of the resonances masses
*The forth column* contains the experimental values of the resonance masses (including errors)

In Fig. 11 and Table 3 one can see a very good agreement between the calculation by formula (1) and the experimental data in the whole range of the $\pi^+\pi^-$ - masses: from threshold (280 MeV/c$^2$) up to the mass of 2400 MeV/c$^2$.

**Therefore it is possible to conclude that resonance of number n contains a whole series of the cluster excitations whose effect is summed and gives total excitation of the scalar field.**

## 6. Discussion and conclusion

We have observed the series of resonances in the system $\pi^+\pi^-$ mesons from the reaction $np \to np\pi^+\pi^-$ at $P_n = 5,20 \pm 0,12 GeV/c$ that have the quantum numbers of $\sigma_0$ – meson ($0^+(0^{++})$) and the masses in the range of $M \leq 1200 MeV/c^2$. The data about these resonances are practically absent.

The widths of the observed peculiarities are sufficiently small. It strongly differs from the experimental data processed by PWA that gives the widths equal to some hundreds MeV/c$^2$.

We could not observe such wide resonances in our experiment.

The statistical significance is sizeable (≥ 5 S.D.) due to the used criterion $\cos\Theta_p^* > 0$ and selection of the events for the corresponding (for each effect) ranges of $X^*_{\pi^+\pi^-}$.

The resonances are evidently produced in the scattering of the $\pi^+$ on $\pi^-$ - mesons. And at the same time it is accompanied by well-ordered excitation of scalar fields.

The study of $\sigma_0$ – mesons in the hot and dense matter will give much more information about the properties of this matter. Therefore the study of the $\sigma_0$ –mesons is extremely important both for the NICA/MPD project and experiments with nuclear beams on all accelerators of the world.

We have not observed the peak of the $f_0$ – resonance at the mass of $M = 980 MeV/c^2$ decay – it was more likely a hole in the distribution of the $\pi^+\pi^-$ - masses in this region. It leads to an idea that $f_0(980)$ – resonance is masked in our experiment by a threshold effect of the $K^+K^-$ - mesons pair production.

**Acknowledgement**

We thank prof. V.L. Lyuboshitz - for the persistent help during discussions of physical results, prof. A.I. Malakhov and prof. S. Vokal - for the interest to this work and presentation of the obtained results at different conferences, dr. M.V.Tokarev - for the fruitful discussions, dr. P.Palazzi - for the interest to our work.